\documentstyle[art12]{article}

\title{Functional representation of the Ablowitz-Ladik hierarchy.}
\author{V.E.Vekslerchik \\
\\
\normalsize\it
  Institute for Radiophysics and Electronics,  \\
\normalsize\it
  National Academy of Sciences of Ukraine,     \\
\normalsize\it
  Proscura Street 12, Kharkov 310085, Ukraine.}

\date{\today}

\begin{document}

\maketitle

\begin{abstract}
The Ablowitz-Ladik hierarchy (ALH) is considered in the framework of the
inverse scattering approach. After establishing the structure of solutions
of the auxiliary linear problems, the ALH, which has been originally
introduced as an infinite system of difference-differential equations is
presented as a finite system of difference-functional equations. The
representation obtained, when rewritten in terms of Hirota's bilinear
formalism, is used to demonstrate relations between the ALH and some other
integrable systems, the Kadomtsev-Petviashvili hierarchy in particular.
\end{abstract}

\section{Introduction.}

Among various concepts of the theory of integrable nonlinear systems
one of the most fruitful is a viewpoint when each integrable equation is
considered as member of an infinite number of related equations ---
hierarchy \cite{N}. So, e.g., the famous nonlinear Schrodinger
equation is the simplest equation of the AKNS hierarchy, its discrete
integrable analog is a member of the Ablowitz-Ladik hierarchy (ALH), etc.
A distinguishing feature of integrable hierarchies is that corresponding
flows commute, i.e., all its equations are compatible. This enables to
consider a hierarchy as one system of equations, i.e., to consider an
infinite number of, say, (1+1)-dimensional partial differential equations
(PDE) (as, for example, in the case of the AKNS hierarchy) or
difference-differential equations (DDE) (in the case of the ALH) as
a (1+$\infty$)-dimensional problem for functions depending on an infinite
number of variables. Such an approach has been intensively studied for
almost all 'classical' integrable equations and has been shown to be a
rather powerful tool for tackling integrable nonlinear problems. A logical
continuation of this method is to 'convert' this infinite number of PDE's
or DDE's into one (or several) functional equation which can be viewed as
a generating function  for equations of the hierarchy: expanding this
functional relation in power series in some auxiliary parameter one can
obtain all equations of the hierarchy.
Such functional equations naturally appear in the Sato theory of soliton
equations \cite{Sato}. Also, this question, especially in the case of the
Kadomtsev-Petviashvili (KP) hierarchy, has been discussed in connection
with the problem of characterization of the Jacobi varietes (see, e.g.,
\cite{ASvM,Shiota} and references therein). Some recent examples of the
functional representation of integrable systems one can find, for
example, in \cite{TT,BK,DickeyStrampp,CarrollKodama} (KP and
dispersionless KP hierarchies) and \cite{UT,Kakei} (Toda hierarchy).

In the present paper I will consider the case of the ALH. After
outlining some basic facts related to the inverse scattering transform
(IST) (section \ref{sec-ALH}) and discussing more comprehensively
corresponding linear problems (section \ref{sec-LP}) I will obtain the
functional representation of the ALH (section \ref{sec-Main}). In the
section \ref{sec-Hirota} I will rewrite the results obtained in terms of
Hirota's bilinear operators. This will expose some relations between the
ALH and other integrable hierarchies, KP hierarchy in particular.

\section{Zero curvature representation of the ALH.} \label{sec-ALH}

The ALH is an infinite
set of ordinal differential-difference equations, that has been introduced
by Ablowitz and Ladik in 1975 \cite{AL1}. The most well-known of these
equations is the discrete nonlinear Schrodinger equation (DNLSE)

\begin{equation}
i \dot q_{n} =
q_{n+1} - 2q_{n} + q_{n-1} -
q_{n}r_{n} \left( q_{n+1} + q_{n-1} \right)
\label{dnlse}
\end{equation}
and the discrete modified KdV equation (DMKdV),

\begin{equation}
\dot q_{n} =
p_{n} \left( q_{n+1} - q_{n-1} \right)
\label{dmkdv}
\end{equation}
where
\begin{equation}
p_{n} = 1 - q_{n}r_{n}, \quad
r_{n} = -\kappa \bar q_{n}, \quad
\kappa = \pm 1
\end{equation}
(see, e.g., \cite{AS}).
All equations of the ALH can be presented as the compatibility condition
for the linear system

\begin{eqnarray}
\Psi_{n+1} &=& U_{n} \Psi_{n}
\label{zcr-sp}\\
\partial_{t} \Psi_{n} &=& V_{n} \Psi_{n}
\label{zcr-evol}
\end{eqnarray}
where $\partial_{t}$ stands for $\partial/\partial t$,
which leads to their zero-curvature representation (ZCR):

\begin{equation}
\partial_{t} U_{n} = V_{n+1} U_{n} - U_{n} V_{n}
\label{ZCR}
\end{equation}
In the standard IST approach developed in \cite{AL1} the matrix $U_{n}$
for the ALH is given by

\begin{equation}
U_{n} =
\pmatrix{ \lambda & r_{n} \cr q_{n} & \lambda^{-1} }
\end{equation}
where $\lambda$ is auxiliary constant parameter. For the elements of the
matrix $V_{n}$,

\begin{equation}
V_{n} = \pmatrix{ a_{n} & b_{n} \cr c_{n} & d_{n} }
\label{abcd}
\end{equation}
one can then obtain from (\ref{ZCR}), the system of equations

\begin{eqnarray}
&&
\lambda \left( a_{n+1} - a_{n} \right) =
   -q_{n} b_{n+1} + r_{n} c_{n}
\label{V-eqs-1}
\\
&&
\lambda^{-1} \left( d_{n+1} - d_{n} \right) =
   q_{n} b_{n} - r_{n} c_{n+1}
\\
&&
\partial_{t} q_{n} =
    q_{n} \left( d_{n+1} - a_{n} \right)
   + \lambda c_{n+1} - \lambda^{-1} c_{n}
\\
&&
\partial_{t} r_{n} =
   r_{n} \left( a_{n+1} - d_{n} \right)
   - \lambda b_{n} + \lambda^{-1} b_{n+1}
\label{V-eqs-4}
\end{eqnarray}
According to \cite{AL1}, they can be chosen as Laurent polynomials in
$\lambda$ in such a way that (\ref{V-eqs-1})-(\ref{V-eqs-4}) hold
automatically for all $\lambda$'s provided $q_{n}$'s, $r_{n}$'s satisfy
some differential relations. It should be noted that one can obtain an
infinite number of the matrices $V_{n}$ (which are Laurent polynomials of
different order) which leads to the infinite number of differential
equations $\partial q_{n} / \partial t = F^{l}_{n}$, ($l=1,2,...$).
According to the widely accepted now viewpoint, as was mentioned in the
Introduction, one can consider $q_{n}$'s and $r_{n}$'s as depending on the
infinite number of 'times', $q_{n} = q_{n}(t_{1}, t_{2}, ...)$ and
consider the $l$th equation of the ALH as describing the flow with respect
to the $l$th variable, $\partial q_{n} / \partial t_{l} = F^{l}_{n}$. I
will also adhere this conception of $q_{n}$'s being functions of the
infinite number of variables, but my approach will slightly differ from
the classical one in the following aspect.  Traditionally it is implied
that all 'times' $t_{l}$ are real, which is grounded from the standpoint
of physical applications, and also is convenient in the framework of the
inverse scattering technique.  I will use instead of real 'times' $t_{l}$
some complex variables $z_{j}, \bar z_{j} \; (j=1,2,...)$, which, as will
shown below, exhibit in a more transparent way some intrinsic properties
of the ALH.  A simple analysis yields that the family of possible
solutions of the system (\ref{V-eqs-1})-(\ref{V-eqs-4}) (and hence the
equations of the hierarchy) can be divided in two subsystems.  One of them
consist of $V$-matrices which are polynomials in $\lambda^{-1}$ (I will
term the corresponding equations as a 'positive' part of hierarchy), and
the other consist of matrices which are polynomials in $\lambda$
('negative' subhierarchy), while in the standard, 'real-time' approach all
the $V$-matrices contain terms proportional to $\lambda^{m}$ together with
the terms proportional to $\lambda^{-m}$ ($m \geq 0$).  Let us consider
first the 'positive' case.  An infinite number of polynomial in
$1/\lambda$ solutions $V_{n}^{j}$ ($j=1,2,...$) possesses the following
structure:

\begin{equation}
V_{n}^{j} = \lambda^{-2} V_{n}^{j-1} +
\pmatrix{ \lambda^{-2} \alpha_{n}^{j} & \lambda^{-1} \beta_{n}^{j} \cr
          \lambda^{-1} \gamma_{n}^{j} & \delta_{n}^{j} }
\end{equation}
where the elements $\alpha_{n}^{j}, ... , \delta_{n}^{j}$ satisfy the
equations

\begin{eqnarray}
&&
\alpha_{n+1}^{j} - \alpha_{n}^{j} =
   - q_{n} \beta_{n+1}^{j} + r_{n} \gamma_{n}^{j}
\\&&
\delta_{n+1}^{j} - \delta_{n}^{j} =
   q_{n} \beta_{n}^{j} - r_{n} \gamma_{n+1}^{j}
\\&&
\partial_{j} q_{n} =
  q_{n} \delta_{n+1}^{j} + \gamma_{n+1}^{j} =
  q_{n} \alpha_{n}^{j+1} + \gamma_{n}^{j+1}
\\&&
\partial_{j} r_{n} =
  - r_{n} \delta _{n}^{j} - \beta_{n}^{j} =
  - r_{n} \alpha_{n+1}^{j+1} - \beta_{n+1}^{j+1}
\end{eqnarray}
with $\partial_{j} = \partial / \partial z_{j}$.
Rewriting this system as

\begin{eqnarray}
&&
\alpha_{n}^{j} = - \delta_{n}^{j-1}
\\&&
\beta_{n}^{j} = \beta_{n-1}^{j-1}
  + r_{n-1} \left( \delta_{n-1}^{j-1} + \delta_{n}^{j-1} \right)
\\&&
\gamma_{n}^{j} = \gamma_{n+1}^{j-1}
  + q_{n} \left( \delta_{n}^{j-1} + \delta_{n+1}^{j-1} \right)
\\&&
\delta_{n}^{j} - \delta_{n+1}^{j} =
  - q_{n} \beta_{n}^{j} + r_{n} \gamma_{n+1}^{j}
\end{eqnarray}
and choosing

\begin{equation}
a_{n}^{0} = 0
\qquad
b_{n}^{0} = 0
\qquad
c_{n}^{0} = 0
\qquad
d_{n}^{0} = - i
\end{equation}
we can obtain consequently

\begin{equation}
\begin{array}{l}
\alpha_{n}^{1} = 0
\cr
\beta_{n}^{1} = -i r_{n-1}
\cr
\gamma_{n}^{1} = -i q_{n}
\cr
\delta_{n}^{1} = i r_{n-1}q_{n}
\end{array}
\qquad
\begin{array}{l}
\alpha_{n}^{2} = - i r_{n-1}q_{n}
\cr
\beta_{n}^{2} = - i r_{n-2}p_{n-1} + i r^{2}_{n-1}q_{n}
\cr
\gamma_{n}^{2} = - i p_{n}q_{n+1} + i r_{n-1}q^{2}_{n}
\cr
\delta_{n}^{2} =
  i r_{n-2}p_{n-1}q_{n} + i r_{n-1}p_{n}q_{n+1} - i r^{2}_{n-1}q^{2}_{n}
\end{array}
\end{equation}
and, in principle, all other matrices $V_{n}^{j}$. This leads to the
infinite system of equations for $q_{n}$, $r_{n}$ some first of which are

\begin{eqnarray}
\partial_{1} q_{n} &=& -ip_{n}q_{n+1}
\label{1}
\\
\partial_{1} r_{n} &=& ir_{n-1}p_{n}
\label{1r}
\\
\cr
\partial_{2} q_{n} &=&
 ir_{n-1}p_{n}q_{n}q_{n+1}+ip_{n}r_{n}q_{n+1}^{2}-ip_{n}p_{n+1}q_{n+2}
\label{2}
\\
\partial_{2} r_{n} &=&
 ir_{n-2}p_{n-1}p_{n}-ir_{n-1}^{2}p_{n}q_{n}-ir_{n-1}p_{n}r_{n}q_{n+1}
\label{2r}
\end{eqnarray}

Analogously, looking for the V-matrices of the form

\begin{equation}
V_{n}^{-j} = \lambda^{2} V_{n}^{-j+1} +
\pmatrix{ \alpha_{n}^{-j} & \lambda \beta_{n}^{-j} \cr
          \lambda \gamma_{n}^{-j} & \lambda^{2} \delta_{n}^{-j} }
\end{equation}
and repeating the procedure described above one can obtain the 'negative'
part of the ALH. Some first of its equations are

\begin{eqnarray}
\partial_{-1} q_{n} &=& -iq_{n-1}p_{n}
\label{-1}
\\
\partial_{-1} r_{n} &=& i p_{n}r_{n+1}
\\
\cr
\partial_{-2} q_{n} &=&
-iq_{n-2}p_{n-1}p_{n}+iq_{n-1}p_{n}q_{n}r_{n+1}+iq_{n-1}^{2}p_{n}r_{n}
\label{-2}
\\
\partial_{-2} r_{n} &=&
-iq_{n-1}p_{n}r_{n}r_{n+1}-ip_{n}q_{n}r_{n+1}^{2}+ip_{n}p_{n+1}r_{n+2}
\end{eqnarray}
where $\partial_{-j} = \partial / \partial \bar z_{j}$
and the overbar denotes the complex conjugation.

Before proceeding further I would like to note that the simplest equations
of the ALH, (\ref{1}) and (\ref{-1}), when rewritten in terms of the real
variables $x = {\rm Re} \, z_{1}$ and $y = {\rm Im} \, z_{1}$ become
exactly the DNLSE (\ref{dnlse}) modified by the substitution
$q_{n} \rightarrow q_{n}\exp(2ix)$ and the DMKdV (\ref{dmkdv})

All equations of the ALH, as well as all equations of other integrable
hierarchies, can be presented in the bilinear form using Hirota's
operators

\begin{equation}
D_{x}^{a} ... D_{y}^{b} \; u \cdot v =
\left.
\left(
  {\partial \over \partial x} - {\partial \over \partial x^{'}}
\right)^{a}
...
\left(
  {\partial \over \partial y} - {\partial \over \partial y^{'}}
\right)^{b}
u(x, y, ...) v(x^{'}, y^{'}, ... )
\right|_{x^{'}=x, \, y^{'}=y, ... }
\end{equation}
To this end consider the functions $\tau_{n}$, $\sigma_{n}$ and
$\rho_{n}$ defined by

\begin{equation}
p_{n} = { \tau_{n-1} \tau_{n+1} \over \tau_{n}^{2} }
\qquad
q_{n} = { \sigma_{n} \over \tau_{n} }
\qquad
r_{n} = { \rho_{n} \over \tau_{n} }
\label{tau-def}
\end{equation}
The first equations of the ALH, (\ref{1}) and (\ref{-1}), can be rewritten
then, using the designation

\begin{equation}
D_{j} = D_{\textstyle z_{j}},
\qquad
\bar D_{j} = D_{\textstyle \bar z_{j}}
\end{equation}
as

\begin{eqnarray}
&&
D_{1} \; \sigma_{n} \cdot \tau_{n} = -i \sigma_{n+1} \tau_{n-1}
\\&&
\bar D_{1} \; \sigma_{n} \cdot \tau_{n} = -i \sigma_{n-1} \tau_{n+1}
\end{eqnarray}
(the corresponding equations for the functions $\rho_{n}$ one can obtain
from these equations using the involution
$\sigma_{n}=-\kappa\bar\rho_{n}$).  The next pair of equations of the ALH,
(\ref{2}) and (\ref{-2}), can be presented as

\begin{eqnarray}
&&
D_{2} \; \sigma_{n}\cdot\tau_{n} = D_{1} \; \sigma_{n+1}\cdot\tau_{n-1}
\label{2qbi}
\\&&
D_{1} \; \tau_{n+1}\cdot\tau_{n} = i \sigma_{n+1}\rho_{n}
\label{2pbi}
\end{eqnarray}
and
\begin{eqnarray}
&&
\bar D_{2}\;\sigma_{n}\cdot\tau_{n}=\bar D_{1}\;\sigma_{n-1}\cdot\tau_{n+1}
\label{-2qbi}
\\&&
\bar D_{1} \; \tau_{n+1}\cdot\tau_{n} = -i \sigma_{n}\rho_{n+1}
\label{-2pbi}
\end{eqnarray}

The bilinear representation of the higher equations of the hierarchy will
be discussed below, and here I would like to mention only one remarkable
fact. By simple calculations one can obtain an alternative representation
for the equations (\ref{2qbi}), (\ref{2pbi}):

\begin{equation}
\left( iD_{2} +D_{11} \right) \; \sigma_{n} \cdot \tau_{n} = 0
\label{1bi-evol}
\end{equation}
(hereafter I will write $D_{x ... y}$ instead of $D_{x} ... D_{y}$)
which involves functions for only one value of the index $n$. In other
words we have presented the differential-difference equations
(\ref{2qbi}), (\ref{2pbi}) as a partial differential equation. One can
rewrite analogously also the equations (\ref{-2qbi}), (\ref{-2pbi}) as
well as all other equations of the ALH. It is a manifestation of the fact
that both 'positive' and 'negative' subhierarchies can be transformed into
hierarchies of (1+1)-dimensional evolution equations for $q=q_{n}$ and
$r=r_{n}$ as functions of $z=z_{1}$ and $z_{j}, \; j=2,3,...$.  Indeed,
expressing from (\ref{1}) and (\ref{1r}) $q_{n+1}$, $q_{n+2}$ and
$r_{n-1}$ in terms of $q$, $q_{z}$, $q_{zz}$, $r$, $r_{z}$ (here the
subscripts indicate derivatives with respect to $z$) one can rewrite the
equations (\ref{2}), (\ref{2r})  as

\begin{eqnarray}
&&
i\partial_{2} q +  q_{zz} + {2 q q_{z} r_{z} \over 1-qr} = 0
\\
&&
-i \partial_{2} r + r_{zz} + {2 r r_{z} q_{z} \over 1-qr} = 0
\end{eqnarray}
and all higher equations of the 'positive' hierarchy in a similar way.
Analogous procedure can surely be performed also for the 'negative' part
of the hierarchy. Some general formulae for such a representation of the
ALH will be obtained in the section \ref{sec-Hirota}.

\section{Solutions of the auxiliary problems.} \label{sec-LP}

Now I an going to construct some solutions of the linear problems
(\ref{zcr-sp}) and (\ref{zcr-evol}) which I rewrite now as

\begin{equation}
\partial_{j} \Psi_{n} = V_{n}^{j} \Psi_{n}
\label{evol}
\end{equation}

The question of solving (\ref{zcr-sp}) and (\ref{evol}) is not the main
one of the present paper, and I discuss it for an illustrative purpose, to
show a way how one can 'deduce' the functional representation of the ALH
which we are looking for, and which can be then proved independently,
without invoking results of this section. That is why I will write down
some results (namely formulae (\ref{psi_plus}) and (\ref{psi_minus}))
without presenting their rigorous proof.

In what follows I will restrict myself to the simplest case

\begin{equation}
\lim_{n \rightarrow \infty} q_{n}, \, r_{n} = 0
\label{zero_BC}
\end{equation}
or, in the $\{ \tau,\sigma,\rho \}$-representation,

\begin{equation}
\lim_{n \rightarrow \infty} \sigma_{n}, \, \rho_{n} = 0,
\qquad
\lim_{n \rightarrow \infty} \tau_{n} = \mbox{constant}
\end{equation}
Presenting the elements of the first column of the matrix $\Psi_{n}$ as

\begin{equation}
\Psi_{n}^{(1)} =
\lambda^{n}
{\tau_{n} \over \tau_{n-1}}
\pmatrix{ \varphi_{n} \cr - \lambda\psi_{n} }
\label{Psi1}
\end{equation}
one can obtain from (\ref{zcr-sp}) the following equations for the
quantities $\varphi_{n}$, $\psi_{n}$:

\begin{eqnarray}
&&
p_{n}\varphi_{n+1} = \varphi_{n} - r_{n}\psi_{n}
\label{SP_pp1}
\\
&&
- \lambda^{2} p_{n}\psi_{n+1} = q_{n}\varphi_{n} - \psi_{n}
\label{SP_pp2}
\end{eqnarray}
which I will solve under the boundary conditions

\begin{equation}
\lim_{n \rightarrow \infty} \varphi_{n} = 1,
\qquad
\lim_{n \rightarrow \infty} \psi_{n} = 0
\label{BC_pp}
\end{equation}
This problem admit solution that can be presented as power series in
$\lambda^{2}$:

\begin{equation}
\pmatrix{ \varphi_{n} \cr \psi_{n} } =
\sum\limits_{m=0}^{\infty} \lambda^{2m}
\pmatrix{ \varphi_{n}^{m} \cr \psi_{n}^{m} }
\end{equation}
Substituting these series in (\ref{SP_pp1}), (\ref{SP_pp2}) one can derive
a system of equations for the quantities $\varphi_{n}^{m}$ and
$\psi_{n}^{m}$, which can be written as follows:

\begin{eqnarray}
&&
\varphi_{n}^{m} - \varphi_{n-1}^{m} = - r_{n-1} \psi_{n}^{m-1}
\label{SP_iter1}
\\
&&
\psi_{n}^{m} = q_{n} \varphi_{n}^{m} + p_{n} \psi_{n+1}^{m-1}
\label{SP_iter2}
\end{eqnarray}
From this system and (\ref{BC_pp}) one can obtain

\begin{equation}
\varphi_{n}^{0} = 1, \qquad \psi_{n}^{0} = q_{n}
\end{equation}
Using the identity

\begin{equation}
\partial_{1} \ln { \tau_{n} \over \tau_{n-1} } = ir_{n-1}q_{n}
\end{equation}
which follows from (\ref{2pbi}) one can perform next iteration:

\begin{equation}
\varphi_{n}^{1} =  {i \over \tau_{n} } \partial_{1} \tau_{n},
\qquad
\psi_{n}^{1} = {i \over \tau_{n} } \partial_{1} \sigma_{n}
\end{equation}
Further, using

\begin{equation}
\partial_{2} \ln { \tau_{n} \over \tau_{n-1} } =
ir_{n-2}p_{n-1}q_{n} + ir_{n-1}p_{n}q_{n+1} - ir_{n-1}^{2}q_{n}^{2}
\end{equation}
one can obtain

\begin{equation}
\varphi_{n}^{2} = { 1 \over 2\tau_{n} }
  \left( i \partial_{2} - \partial_{11} \right) \tau_{n}
\qquad
\psi_{n}^{2} = { 1 \over 2\tau_{n} }
  \left( i \partial_{2} - \partial_{11} \right) \sigma_{n}
\end{equation}
Iterating further the system (\ref{SP_iter1}), (\ref{SP_iter2}) one can
conclude that the quantities $\tau_{n}\varphi_{n}^{m}$ and
$\tau_{n}\psi_{n}^{m}$ are the coefficients of the Taylor expansion for
the functions $\tau_{n}(z_{1} + i\lambda^{2}, z_{2} + i\lambda^{4}/2,
...)$ and $\sigma_{n}(z_{1} + i\lambda^{2}, z_{2} + i\lambda^{4}/2, ...)$.
Moreover, it can be shown that the column (\ref{Psi1}) with

\begin{equation}
\varphi_{n} =
  { \tau_{n}( z_{k} + i \lambda^{2k}/k, \bar z_{k})
    \over \tau_{n}( z_{k} , \bar z_{k}) },
\qquad
\psi_{n} =
  { \sigma_{n}( z_{k} + i \lambda^{2k}/k, \bar z_{k})
    \over \tau_{n}( z_{k}, \bar z_{k} )}
\label{sls_LP}
\end{equation}
where $\tau_{n}$ and $\sigma_{n}$ are solutions of the 'positive'
subhierarchy, solve the linear problems (\ref{evol}) for $j=1,2,...$.
Here the designation

\begin{equation}
f(z_{k}, \bar z_{k})
\equiv
f(z_{1}, z_{2}, ... \; \bar z_{1}, \bar z_{2}, ...)
\end{equation}
is used.

Considering in a similar way the second column of the matrix $\Psi_{n}$
one can obtain the following matrix solution for the linear problems of
the 'positive' subhierarchy (i.e the problems (\ref{zcr-sp}) and
(\ref{evol}) for $j=1,2,...$):

\begin{equation}
\Psi_{n}^{+} =
{ 1 \over \tau_{n-1}(z_{k}, \bar z_{k}) }
\pmatrix{
  \lambda^{n}
  \tau_{n}(z_{k} + i \lambda^{2k} / k, \bar z_{k})
&
  \lambda^{-n+1} \exp(- i \phi)
  \rho_{n-1}(z_{k} - i \lambda^{2k} / k, \bar z_{k})
\cr\cr
  - \lambda^{n+1}
  \sigma_{n}(z_{k} + i \lambda^{2k} / k, \bar z_{k})
&
  \lambda^{-n}\exp(- i \phi)
  \tau_{n-1}(z_{k} - i \lambda^{2k} / k, \bar z_{k})
}
\label{psi_plus}
\end{equation}
where

\begin{equation}
\phi = \sum_{k=1}^{\infty} \lambda^{-2k} z_{k}
\end{equation}
Analogously, for the linear problems  of the 'negative'
subhierarchy, (\ref{zcr-sp}) and (\ref{evol}) for $j=-1,-2,...$, one
can obtain solution

\begin{equation}
\Psi_{n}^{-} =
{ 1 \over \tau_{n-1}(z_{k}, \bar z_{k}) }
\pmatrix{
  \lambda^{n} \exp(i \tilde\phi)
  \tau_{n-1}(z_{k}, \bar z_{k} + i \lambda^{-2k} / k)
&
  - \lambda^{-n-1}
  \rho_{n}(z_{k}, \bar z_{k} - i \lambda^{-2k} / k)
\cr\cr
  \lambda^{n-1} \exp(i \tilde\phi)
  \sigma_{n-1}(z_{k}, \bar z_{k} + i \lambda^{-2k} / k)
&
  \lambda^{-n}
  \tau_{n}(z_{k}, \bar z_{k} - i \lambda^{-2k} / k)
}
\label{psi_minus}
\end{equation}
where

\begin{equation}
\widetilde\phi = \sum_{k=1}^{\infty} \lambda^{2k} \bar z_{k}
\end{equation}

The obtained formulae (\ref{psi_plus}), (\ref{psi_minus}) are one more
illustration for the fact that an integrable hierarchy is more than a
collection of solvable equations, and considering an hierarchy one can
sometimes obtain more 'transparent' results than if dealing with one
particular equation. Such approach (and such results) is not entirely new,
it had been applied earlier to other hierarchies, say AKNS \cite{N},
though, to my knowledge, for the ALH this has been done for the first time
in this work.

\section{The main result.}            \label{sec-Main}

In the previous section we constructed the matrices $\Psi_{n}^{+}$
($\Psi_{n}^{-}$) which are solutions of the discrete auxiliary problem
(\ref{zcr-sp}) and the 'positive' ('negative') set of evolution linear
problems (\ref{evol}). And though validity of these results should be
discussed more accurately, they give us sufficient hint to obtain the main
result of this work, namely the functional representation of the ALH. The
matrix equation $\Psi_{n+1}^{+} = U_{n}\Psi_{n}^{+}$ after some
transformations can be rewritten in the following way:

\begin{eqnarray}
&&
\sigma_{n}(z_{k} + i\lambda^{2k}/k, \bar z_{k})
\tau_{n}(z_{k}, \bar z_{k})
-
\sigma_{n}(z_{k}, \bar z_{k})
\tau_{n}(z_{k} + i\lambda^{2k}/k, \bar z_{k}) =
\label{premain-q}
\\&& \qquad\qquad =
\lambda^{2} \;
\tau_{n-1}(z_{k}, \bar z_{k})
\sigma_{n+1}(z_{k} + i\lambda^{2k}/k, \bar z_{k})
\cr\cr
&&
\rho_{n}(z_{k}, \bar z_{k})
\tau_{n}(z_{k} + i\lambda^{2k}/k, \bar z_{k})
-
\rho_{n}(z_{k} + i\lambda^{2k}/k, \bar z_{k})
\tau_{n}(z_{k}, \bar z_{k}) =
\label{premain-r}
\\&& \qquad\qquad =
\lambda^{2} \;
\rho_{n-1}(z_{k}, \bar z_{k})
\tau_{n+1}(z_{k} + i\lambda^{2k}/k, \bar z_{k})
\cr\cr
&&
\tau_{n}(z_{k} + i\lambda^{2k}/k, \bar z_{k})
\tau_{n}(z_{k}, \bar z_{k})
-
\tau_{n-1}(z_{k}, \bar z_{k})
\tau_{n+1}(z_{k} + i\lambda^{2k}/k, \bar z_{k}) =
\label{premain-p}
\\&& \qquad\qquad =
\sigma_{n}(z_{k} + i\lambda^{2k}/k, \bar z_{k})
\rho_{n}(z_{k}, \bar z_{k})
\nonumber
\end{eqnarray}
Now we can forget about were these equations originate from and consider
them as a system of three difference-functional {\it equations} for
unknown functions $\tau_{n}$, $\sigma_{n}$ and $\rho_{n}$. This system is
compatible with all 'positive' flows of the ALH: if $\tau_{n}$,
$\sigma_{n}$ and $\rho_{n}$ solve (\ref{premain-q}) - (\ref{premain-p})
then $\Psi_{n}^{+}$ satisfy all equations (\ref{evol}) for $j=1,2,...$. To
prove this consider the quantities

\begin{eqnarray}
X_{n}^{j} &=&
  \partial_{j}\varphi_{n}
  - a_{n}^{j} \varphi_{n} + \lambda b_{n}^{j} \psi_{n}
\\
Y_{n}^{j} &=&
  \partial_{j}\psi_{n}
  +\lambda^{-1} c_{n}^{j} \varphi_{n} -  d_{n}^{j} \psi_{n}
\end{eqnarray}
where $\varphi_{n}$, $\psi_{n}$ are defined by

\begin{equation}
\varphi_{n} =
  { \tau_{n}( z_{k} + i \lambda^{2k}/k, \bar z_{k})
    \over \tau_{n-1}( z_{k} ), \bar z_{k}},
\qquad
\psi_{n} =
  { \sigma_{n}( z_{k} + i \lambda^{2k}/k, \bar z_{k})
    \over \tau_{n-1}( z_{k}, \bar z_{k} )}
\label{phi_psi}
\end{equation}
(note that these functions $\varphi_{n}$, $\psi_{n}$  differ from ones
defined by (\ref{sls_LP}) in the factor $\tau_{n}/\tau_{n-1}$) and
$a_{n}^{j}, ..., d_{n}^{j}$ are elements of the matrix $V_{n}^{j}$ (see
(\ref{abcd})).  Using the identities

\begin{eqnarray}
&&
\varphi_{n+1} - \varphi_{n} + r_{n} \psi_{n} = 0
\label{varphi_rec}
\\&&
\lambda^{2} \psi_{n+1} - \psi_{n} + q_{n}\varphi_{n} = 0
\label{psi_rec}
\end{eqnarray}
which follow from (\ref{premain-q}), (\ref{premain-p}) one can
straightforwardly verify the fact that the combination
$X_{n+1}-X_{n}+r_{n}Y_{n}$ can be presented as

\begin{eqnarray}
X_{n+1} - X_{n} + r_{n} Y_{n} &=&
\partial_{j} \left[\varphi_{n+1}-\varphi_{n}+r_{n}\psi_{n}\right] +
\\&&
+ \left[
  a_{n}^{j} - a_{n+1}^{j} -
  \lambda^{-1} q_{n} b_{n+1}^{j} + \lambda^{-1} r_{n} c_{n}^{j}
\right] \varphi_{n}
\cr&&
+ \left[
  - \partial_{j} r_{n}
  + \left( a_{n+1}^{j} - d_{n}^{j} \right) -
  \lambda b_{n}^{j} + \lambda^{-1} b_{n+1}^{j}
\right] \psi_{n}
\nonumber
\end{eqnarray}
It can be easily seen from (\ref{varphi_rec}) together
with (\ref{V-eqs-1}) and (\ref{V-eqs-4}) that all expressions in square
brackets are equal to zero, i.e.

\begin{equation}
X_{n+1} - X_{n} + r_{n} Y_{n} = 0
\label{X_rec}
\end{equation}
Analogously, calculating in a similar way
$\lambda^{2} Y_{n+1} - Y_{n} + q_{n} X_{n}$ one can obtain

\begin{equation}
\lambda^{2} Y_{n+1} - Y_{n} + q_{n} X_{n} = 0
\label{Y_rec}
\end{equation}
Presenting $X_{n}^{j}$ and $Y_{n}^{j}$ as

\begin{equation}
X_{n}^{j} =
  { \tau_{n} \over \tau_{n-1} }
  \sum_{m=0}^{\infty} \lambda^{2m} X_{n,m}^{j}
\qquad
Y_{n}^{j} =
  { \tau_{n} \over \tau_{n-1} }
  \sum_{m=0}^{\infty} \lambda^{2m} Y_{n,m}^{j}
\end{equation}
one can derive the the following recurrence for the coefficients
$X_{n,m}^{j}$, $Y_{n,m}^{j}$:

\begin{eqnarray}
&&
X_{n,m}^{j} - X_{n-1,m}^{j} = - {\rho_{n-1} \over \tau_{n}} Y_{n,m-1}^{j}
\label{Xm_rec}
\\&&
Y_{n,m}^{j} = p_{n} Y_{n+1,m-1}^{j} + q_{n} X_{n,m}^{j}
\label{Ym_rec}
\end{eqnarray}
It can be shown that $X_{n}(\lambda=0) = Y_{n}(\lambda=0) = 0$, i.e.
$X_{n,0}^{j} = Y_{n,0}^{j} = 0$. Then, (\ref{Xm_rec}) yields $X_{n,1}^{j} =
\mbox{constant}$, this constant, as follows from the boundary conditions for
$\varphi_{n}$ and $\psi_{n}$ is zero, $X_{n,1}^{j} = 0$. This, in its
turn, together with (\ref{Ym_rec}) leads to $Y_{n,1}^{j} = 0$, etc.
Repeating iterations one can obtain $X_{n}^{j} = Y_{n}^{j} = 0$, which
implies that the column
$\left( \varphi_{n} \, , \, -\lambda\psi_{n} \right)^{T}$
is a solution of the equation

\begin{equation}
\partial_{j} \pmatrix{ \varphi_{n} \cr - \lambda \psi_{n}} =
V_{n}^{j} \pmatrix{ \varphi_{n} \cr - \lambda \psi_{n}}
\label{evol-1}
\end{equation}
Thus, we have shown that equations (\ref{premain-q}) - (\ref{premain-p}),
which I would like to rewrite in a more symmetrical way using the
substitutions $z_{k} \rightarrow z_{k} \mp i\lambda^{2k}/2k$,

\begin{eqnarray}
&&
\sigma_{n}(z_{k} + i\delta^{k}/2k) \tau_{n}(z_{k} - i\delta^{k}/2k) -
\sigma_{n}(z_{k} - i\delta^{k}/2k) \tau_{n}(z_{k} + i\delta^{k}/2k) =
\label{main-q}
\\&& \qquad\qquad =
\delta \;
\tau_{n-1}(z_{k} - i\delta^{k}/2k) \sigma_{n+1}(z_{k} + i\delta^{k}/2k)
\cr\cr
&&
\rho_{n}(z_{k} - i\delta^{k}/2k) \tau_{n}(z_{k} + i\delta^{k}/2k) -
\rho_{n}(z_{k} + i\delta^{k}/2k) \tau_{n}(z_{k} - i\delta^{k}/2k) =
\label{main-r}
\\&& \qquad\qquad =
\delta \;
\rho_{n-1}(z_{k} - i\delta^{k}/2k) \tau_{n+1}(z_{k} + i\delta^{k}/2k)
\cr\cr
&&
\tau_{n}(z_{k} + i\delta^{k}/2k) \tau_{n}(z_{k} - i\delta^{k}/2k) -
\tau_{n-1}(z_{k} - i\delta^{k}/2k) \tau_{n+1}(z_{k} + i\delta^{k}/2k) =
\label{main-p}
\\&& \qquad\qquad =
\sigma_{n}(z_{k} + i\delta^{k}/2k) \rho_{n}(z_{k} - i\delta^{k}/2k)
\nonumber
\end{eqnarray}
(here $\delta$ is used instead of $\lambda^{2}$ and dependence on the
conjugated coordinates, $\bar z_{k}$, is temporarily omitted) are indeed
compatible with the 'positive' flows (\ref{evol}) for $j=1,2,...$, and one
can consider them as being equivalent to the 'positive' part of the ALH.
Expanding (\ref{main-q})-(\ref{main-p}) in powers of $\delta$
one will obtain an infinite number of differential-difference equation that
can be transformed to the ones from the ALH. So, e.g., the equations which
correspond to the first power of $\delta$

\begin{eqnarray}
&&
\left( \partial_{1} \sigma_{n} \right) \tau_{n} -
\sigma_{n} \left( \partial_{1} \tau_{n} \right) =
-i \tau_{n-1} \sigma_{n+1}
\\&&
\left( \partial_{1} \rho_{n} \right) \tau_{n} -
\rho_{n} \left( \partial_{1} \tau_{n} \right) =
i \rho_{n-1} \tau_{n+1}
\end{eqnarray}
are obviously the equations (\ref{1}), (\ref{1r}) rewritten in the
$\{ \tau_{n}, \sigma_{n}, \rho_{n} \}$-representation. The equations which
correspond to the second power of $\delta$ are equivalent to the second
pair of equations of the ALH, (\ref{2}) and (\ref{2r}), etc.

Analogously, the 'negative' part of the ALH can be written as following
functional equations:

\begin{eqnarray}
&&
\sigma_{n}(\bar z_{k} + i\widetilde\delta^{k}/2k)
\tau_{n}(\bar z_{k} - i\widetilde\delta^{k}/2k) -
\sigma_{n}(\bar z_{k} - i\widetilde\delta^{k}/2k)
\tau_{n}(\bar z_{k} + i\widetilde\delta^{k}/2k) =
\label{main-barq}
\\&& \qquad\qquad =
\widetilde\delta \;
\sigma_{n-1}(\bar z_{k} + i\widetilde\delta^{k}/2k)
\tau_{n+1}(\bar z_{k} - i\widetilde\delta^{k}/2k)
\cr
\cr
&&
\rho_{n}(\bar z_{k} - i\widetilde\delta^{k}/2k)
\tau_{n}(\bar z_{k} + i\widetilde\delta^{k}/2k) -
\rho_{n}(\bar z_{k} + i\widetilde\delta^{k}/2k)
\tau_{n}(\bar z_{k} - i\widetilde\delta^{k}/2k) =
\label{main-barr}
\\&& \qquad\qquad =
\widetilde\delta \;
\tau_{n-1}(\bar z_{k} + i\widetilde\delta^{k}/2k)
\rho_{n+1}(\bar z_{k} - i\widetilde\delta^{k}/2k)
\cr
\cr
&&
\tau_{n}(\bar z_{k} + i\widetilde\delta^{k}/2k)
\tau_{n}(\bar z_{k} - i\widetilde\delta^{k}/2k) -
\tau_{n-1}(\bar z_{k} + i\widetilde\delta^{k}/2k)
\tau_{n+1}(\bar z_{k} - i\widetilde\delta^{k}/2k) =
\label{main-barp}
\\&& \qquad\qquad =
\sigma_{n}(\bar z_{k} + i\widetilde\delta^{k}/2k)
\rho_{n}(\bar z_{k} - i\widetilde\delta^{k}/2k)
\nonumber
\end{eqnarray}
where $\widetilde\delta$ is used instead of $\lambda^{-2}$ and dependence
on $z_{k}$'s is omitted.

Equations (\ref{main-q}) - (\ref{main-p}) and
(\ref{main-barq}) - (\ref{main-barp}) are the main result of this paper.
They present an infinite number of the difference-differential equations
of the ALH under the zero boundary conditions (\ref{zero_BC}) in the form
of six difference-functional equations.
Thus we have derived the 'functional' representation of the ALH.
Analogous results can obtained for
some other classes of boundary conditions, say, for the so-called
finite-density ($q_{n} \to $ constant as $n \to \pm\infty$) or
quasiperiodical ones. Before proceeding further I would like to say few
words on the following question. We have split the ALH into two
subhierarchies (the 'positive' and 'negative' ones) which seems to be
rather natural: one of the subhierarchies can be obtained from the other
using the complex conjugation. Nevertheless, it would be interesting to
obtain, instead of two sets of functional equations ((\ref{main-q}) -
(\ref{main-p}) for the 'positive' hierarchy and (\ref{main-barq}) -
(\ref{main-barp}) for the 'negative' one) one set of equations which takes
into account both 'positive' and 'negative' flows. I cannot do this at
present, and it will be a subject of following studies.

\section{Hirota's representation of the ALH.} \label{sec-Hirota}

It is already known fact that the $D$-operators calculus that has been
invented by Hirota is not only an ingenious tool for deriving some
families of solutions for integrable equations. It is a convenient way of
operating with integrable hierarchies, which enables to reveal some
regularities in their structure. Now I am going to rewrite the main
results in Hirota's formalism, which, in addition, will demonstrate
some interesting features of the ALH. In what follows I will deal only
with 'positive' subhierarchy, because for the 'negative' one all results
can be obtained using the complex conjugation
($\sigma_{n} = -\kappa\bar\rho_{n}$, etc).
Using the following property of the Hirota's operators

\begin{equation}
\exp \left\{ a D_{z} \right\} f(z) \cdot g(z) =
f(z+a) g(z-a)
\end{equation}
and introducing

\begin{equation}
D(\delta) =
\sum_{k=1}^{\infty} { \delta^{k} \over k } D_{k}
\end{equation}
one can rewrite (\ref{main-q})-(\ref{main-p})  as

\begin{eqnarray}
&&
\exp\left[ {i \over 2} D(\delta) \right]
\left(
  \sigma_{n} \cdot \tau_{n} - \tau_{n} \cdot \sigma_{n}
  - \delta \; \sigma_{n+1} \cdot \tau_{n-1}
\right) = 0
\label{Hirota-q}
\\
&&
\exp\left[ {i \over 2} D(\delta) \right]
\left(
\rho_{n} \cdot \tau_{n} -
\tau_{n} \cdot \rho_{n} +
\delta \; \tau_{n+1} \cdot \rho_{n-1}
\right) = 0
\label{Hirota-r}
\\
&&
\exp\left[ {i \over 2} D(\delta) \right]
\left(
\tau_{n+1} \cdot \tau_{n-1} -
\tau_{n} \cdot \tau_{n} +
\sigma_{n} \cdot \rho_{n}
\right) = 0
\label{Hirota-p}
\end{eqnarray}
One of the advantages of this viewpoint is that one can obtain explicit
form of the $j$th equation of the ALH, which hardly can be done in the
framework of the standard IST technique discussed in the section
\ref{sec-ALH}.  This can be done in terms of the Schur's polynomials

\begin{equation}
\exp\left\{ \sum_{m=1}^{\infty} x^{m} f_{m} \right\} =
\sum_{m=0}^{\infty} x^{m} \chi_{m}\left( f_{1}, f_{2}, ... \right)
\end{equation}
(I will use below the designation
$\chi_{m}\left( f_{k} \right) \equiv
\chi_{m}\left( f_{1}, f_{2}, ... \right)$).
By simple calculations equations (\ref{Hirota-q}), (\ref{Hirota-r}) which
can be rewritten as

\begin{eqnarray}
&&
2i \sin\left[ {1 \over 2} D(\delta) \right] \sigma_{n} \cdot \tau_{n} =
  \phantom{-}\delta \;
  \exp\left[ {i \over 2} D(\delta) \right] \sigma_{n+1} \cdot \tau_{n-1}
\\
&&
2i \sin\left[ {1 \over 2} D(\delta) \right] \rho_{n} \cdot \tau_{n} =
  - \delta \;
  \exp\left[ {i \over 2} D(\delta) \right] \tau_{n+1} \cdot \rho_{n-1}
\end{eqnarray}
and equation (\ref{Hirota-p}) can be presented as

\begin{eqnarray}
&&
\left\{
\chi_{j}\left( {i D_{k} \over 2 k} \right) -
\chi_{j}\left( - {i D_{k} \over 2 k} \right)
\right\}
  \sigma_{n} \cdot \tau_{n}  =
\chi_{j-1}\left( {i D_{k} \over 2 k} \right)
  \sigma_{n+1} \cdot \tau_{n-1}
\label{Schur-q}
\\
&&
\left\{
\chi_{j}\left( {i D_{k} \over 2 k} \right) -
\chi_{j}\left( - {i D_{k} \over 2 k} \right)
\right\}
  \rho_{n} \cdot \tau_{n}  =
- \chi_{j-1}\left( {i D_{k} \over 2 k} \right)
  \tau_{n+1} \cdot \rho_{n-1}
\\
&&
\chi_{j}\left( {i D_{k} \over 2 k} \right)
\left(
  \tau_{n+1} \cdot \tau_{n-1} -
  \tau_{n} \cdot \tau_{n} +
  \sigma_{n} \cdot \rho_{n}
\right) = 0
\label{Schur-p}
\end{eqnarray}
for $j=1,2,...$

It has been noticed in section \ref{sec-ALH} that 'positive' subhierarchy
of the ALH (as well as the 'negative' one) can be presented as a hierarchy
of partial differential equations, which can be easily derived from
(\ref{main-q})-(\ref{main-p}). Using the identities

\begin{eqnarray}
&&
D_{1}
 \sigma_{n}(z_{k}+i\delta^{k}/2k)\cdot\tau_{n}(z_{k}-i\delta^{k}/2k) =
-i\sigma_{n+1}(z_{k}+i\delta^{k}/2k)\tau_{n-1}(z_{k}-i\delta^{k}/2k)
\\&&
D_{1}
 \tau_{n}(z_{k}+i\delta^{k}/2k)\cdot\rho_{n}(z_{k}-i\delta^{k}/2k) =
-i\tau_{n+1}(z_{k}+i\delta^{k}/2k)\rho_{n-1}(z_{k}-i\delta^{k}/2k)
\\&&
D_{1}
 \tau_{n}(z_{k}+i\delta^{k}/2k)\cdot\tau_{n}(z_{k}-i\delta^{k}/2k) =
 i \delta
 \sigma_{n+1}(z_{k}+i\delta^{k}/2k)\rho_{n-1}(z_{k}-i\delta^{k}/2k)
\end{eqnarray}
which follow from (\ref{phi_psi}) and (\ref{evol-1}) for $j=1$, equations
(\ref{main-q}) - (\ref{main-p}) can be rewritten as

\begin{equation}
\widehat G(\delta)
\pmatrix{
  \sigma \cdot \tau \cr
  \tau   \cdot \rho \cr
  \sigma \cdot \rho + \tau \cdot \tau} = 0
\label{pos_hie}
\end{equation}
where $\sigma$, $\rho$ and $\tau$ stand for $\sigma_{n}$, $\rho_{n}$ and
$\tau_{n}$ with $n$ being fixed and the operator $\widehat G(\delta)$ is
defined by

\begin{equation}
\widehat G(\delta) =
2 i \sin\left[ {1 \over 2} D(\delta) \right] -
i \delta \; D_{1} \;
\exp\left[ {i \over 2} D(\delta) \right]
\end{equation}
Expanding (\ref{pos_hie}) in power series in $\delta$ one can obtain a
hierarchy of partial differential equations

\begin{equation}
\widehat G_{j}
\pmatrix{
  \sigma \cdot \tau \cr
  \tau   \cdot \rho \cr
  \sigma \cdot \rho + \tau \cdot \tau
} = 0
\qquad
j=2,3,...
\label{pos_pde}
\end{equation}
where operators $\widehat G_{j}$ are defined by

\begin{equation}
\widehat G(\delta) =
\sum_{j=2}^{\infty} { \delta^{j} \over j } \widehat G_{j}
\end{equation}
Some first equations of this hierarchy are ones given by (\ref{pos_pde})
with

\begin{eqnarray}
\widehat G_{2} &=& iD_{2} + D_{11}
\\
\widehat G_{3} &=& iD_{3} + {3 \over 4} D_{21} + {i \over 4} D_{111}
\\
\widehat G_{4} &=& iD_{4} + {2 \over 3} D_{31} + {i \over 4} D_{211}
    - {1 \over 12} D_{1111}
\end{eqnarray}

A rather interesting consequence of (\ref{main-q}) - (\ref{main-p}) can be
obtained by excluding $\sigma_{n}$ and $\rho_{n}$. It can be
straightforwardly shown using (\ref{phi_psi}) and (\ref{evol-1}) for
$j=1,2$ that

\begin{equation}
\left[ 2 D_{1} - \delta \left(D_{2} + i D_{11} \right) \right]
 \tau_{n}(z_{k}+i\delta^{k}/2k)\cdot\tau_{n}(z_{k}-i\delta^{k}/2k) = 0
\end{equation}
or, using again the $\exp\left[ i D(\delta) / 2 \right]$ operator,

\begin{equation}
\left[ 2 D_{1} - \delta \left(D_{2} + i D_{11} \right) \right]
\exp\left[ {i \over 2} D(\delta) \right]
 \tau \cdot \tau = 0
\end{equation}
where $\tau \equiv \tau_{n}$ (for any $n$). Expanding this equation in
powers of $\delta$ one can obtain again an infinite number of equations,
this time for one function, $\tau$. Few first of them are

\begin{eqnarray}
&&
\left( 4 D_{31} - 3 D_{22} + D_{1111} \right) \tau \cdot \tau = 0
\label{KP}
\\
&&
\left( 3 D_{41} - 2 D_{32} + D_{2111} \right) \tau \cdot \tau = 0
\\
&&
\left(
  96 D_{51} - 60 D_{42} + 20 D_{3111} +
  15 D_{2211} - D_{111111}
\right)
\tau \cdot \tau = 0
\end{eqnarray}
It seems to be interesting that equation (\ref{KP}) is nothing other than
the Kadomtsev-Petviashvili equation. Indeed, it can be verified by
straightforward (though rather cumbersome) calculations that the quantity
$u=r_{n-1}p_{n}q_{n+1}$ for any $n$ solves the equation

\begin{equation}
\partial_{1}
\left(
4 \partial_{3} u
+ \partial_{111} u
+ 12 u \, \partial_{1} u
\right) =
3 \partial_{22} u
\end{equation}
So, we have obtained the remarkable
result: the Kadomtsev-Petviashvili equation turns out to be 'embedded' in
the ALH!

\section{Conclusion.}

In the present work it has been obtained a representation of the ALH in
the form of difference-functional equations. This result seems to be
interesting from several viewpoints. First, it clearly demonstrates common
origin of all equations of the hierarchy. Second, such approach can be
useful as an easy tool for generating of big number of solutions for the
ALH, such as, first of all, multisoliton, 'Wronskian' and some other ones.
An interesting transformation of the results obtained to the already known
ones arises when one considers the problem of quasiperiodical solutions.
The functional relations (\ref{main-q}) - (\ref{main-p}) and
(\ref{main-barq}) - (\ref{main-barp}) become the Fay's trisecant formulae
for the $\theta$-functions of Riemann surfaces. Also, bilinear functional
representation of the ALH can provide answers to some more general
questions of the theory of integrability, such as, e.g., classification of
the Hirota's polynomials: note that (\ref{Schur-q}) can be viewed as
explicit expression for the Hirota's polynomials of arbitrary order.

The last moment I would like to discuss here, and which seems to be one of
the most interesting, is the question of, so to say, 'universality' of the
ALH. It is a known fact that some equations cam be 'embedded' into the
ALH. It has been shown that the ALH 'contains' the 2D Toda lattice
\cite{UT} (see also \cite{2dtl}), O(3,1) $\sigma$-model \cite{sigma},
the Davey-Stewartson (DS) equation and the Ishimori model \cite{DS}. In
the last paper it has been shown that the derivative nonlinear Schrodinger
equation can also be 'embedded' into the ALH, which implies that the same
can be done for the AKNS as well. In the present paper I have shown that
the Kadomtsev-Petviashvili equation (hierarchy) can also be composed of
the ALH-flows. The results of the works \cite{UT,2dtl,sigma,DS}
(her I would like to mention also the papers \cite{Levi,SY}) are in some
sense 'empirical' facts: one can easily verify them by simple
calculations, but one can hardly find there an answer on the question why
do such apparently different models turn out to be interrelated. At the
same time the approach described above, and especially the results
presented in the section \ref{sec-Hirota} provide some insight into this
problem.  An impression arises that the ALH is, in some sense, the most
general hierarchy, at least among the 'classical' ones. Surely, such
statements need much more careful grounding, but the fact that the AKNS,
DS, KP hierarchies may can be obtained from the ALH via reductions
$\{\tau_{n},\sigma_{n},\rho_{n}\} \to \{\tau,\sigma,\rho\} \to \{\tau\}$,
to my mind, indicate that the hypothesis of 'universality' of the ALH is
not senseless and worth further studies.


\end{document}